\documentclass[conference]{IEEEtran}
\IEEEoverridecommandlockouts
\usepackage{cite}
\usepackage{amsmath,amssymb,amsfonts}
\usepackage{subfig}
\usepackage{float}
\usepackage{caption}
\usepackage{algpseudocode}
\usepackage{algorithm}
\usepackage{graphicx}
\usepackage{textcomp}
\usepackage{xcolor}
\usepackage{booktabs} 

\def\BibTeX{{\rm B\kern-.05em{\sc i\kern-.025em b}\kern-.08em
    T\kern-.1667em\lower.7ex\hbox{E}\kern-.125emX}}
\begin{document}

\title{Real-Time Minimum-Energy Operating-Point Tracking for Battery-Powered Micro DC Motors Under Dynamically Variable Loading}

\author{
  \IEEEauthorblockN{Tzu-Hsiang Huang\textsuperscript{1}, 
    Haojian Lu\textsuperscript{1}, 
    Hen-Wei Huang\textsuperscript{2,3,†}, 
    Tan Rong\textsuperscript{4,†}}
  \IEEEauthorblockA{
    \textsuperscript{1}College of Control Science and Engineering, Zhejiang University, Hangzhou, China\\
    \textsuperscript{2}School of Electrical and Electronic Engineering, Nanyang Technological University, Republic of Singapore\\
    \textsuperscript{3}Lee Kong Chian School of Medicine, Nanyang Technological University, Republic of Singapore\\
    \textsuperscript{4}Guangdong Provincial Key Lab of Robotics and Intelligent System, \\
    Shenzhen Institutes of Advanced Technology, Chinese Academy of Sciences, Shenzhen, China\\
    Email: 22332159@zju.edu.cn
  }
  \thanks{This work was supported by National Key R\&D Program of China (2025YFE0113300),
    National Natural Science Foundation of China (62303407, T2293720/T2293724),
    Xiaomi Foundation.}
}

\maketitle
%%%%%%%%%%%%%%%%%%%%%%%%%%%%%%%%%%%%%%%%%%%%%%%%%%%%%%%%%%%%%%%%%%%%%%%%%%%%%%%%
\begin{abstract}

Micro DC brushed motors are widely deployed in battery-powered biomedical systems, where limited energy budgets and variable physiological loading impose stringent efficiency and safety constraints. However, conventional actuation strategies rely on conservative voltage margins to avoid stalling, leading to systematic energy inefficiency. Furthermore, existing methods primarily optimize steady-state performance, neglecting the energy required to complete individual actuation cycles under dynamic conditions.
This paper reveals that the energy consumption per mechanical cycle of a DC motor exhibits a non-monotonic dependence on driving voltage, with a load-dependent minimum that shifts with external loading. Based on this insight, we propose a real-time operating-point tracking method that enables the motor to autonomously converge to its minimum-energy condition. A lightweight load metric derived from current waveform features is introduced to detect load variation, and a two-phase adaptive voltage strategy is developed to track the optimal operating point online.
Experimental results demonstrate that the proposed method can track the new minimum-energy operating region under both low-to-high and high-to-low loading transitions. With 3-cycle averaging, the mean response time is 11.55s for the low-to-high case and 11.16s for the high-to-low case, while the mean convergence voltage is 2.73V and 2.0V, respectively.
\end{abstract}

\begin{IEEEkeywords}
energy consumption, load metric, brushed DC motor
\end{IEEEkeywords}

% ---------------------------------------------------------------
% I. INTRODUCTION
% ---------------------------------------------------------------
\section{INTRODUCTION}

% Micro brushed DC motors play an vital role in medical applications, including subcutaneous implants, medical robotics, and micro-electromechanical systems (MEMS). These systems typically rely on miniaturized power sources such as button cell batteries characterized by small volume, limited energy density, and low storage capacity, which poses challenges to the long-term performance and reliability of the system.
% Recent studies on micro dc motors have primarily focused on improving output torque, optimizing mechanical structures, and fault diagnosis based on current signature analysis. The above studies have devoted insufficient attention to the characteristic of MEMS operating under the constraints of miniaturized power sources.

% Micro brushed DC motors are widely used in miniature electromechanical systems, such as medical robots, implantable devices, and other small-scale actuation platforms\cite{Sun2024MicroNanoMedicine, Yan2022ImplantableActuators}. In these applications, the available energy is usually limited by the small size and capacity of the power source\cite{Roundy2003VibrationPowerSource}. Therefore, in addition to achieving the required actuation performance, reducing energy consumption is also important for extending system operating time. This issue becomes more significant in applications where frequent battery replacement or recharging is difficult.

\begin{figure}[t] % 或者 [!htb], [!h] 等，根據您的排版需求調整
  \centering
  \includegraphics[width=0.75\columnwidth]{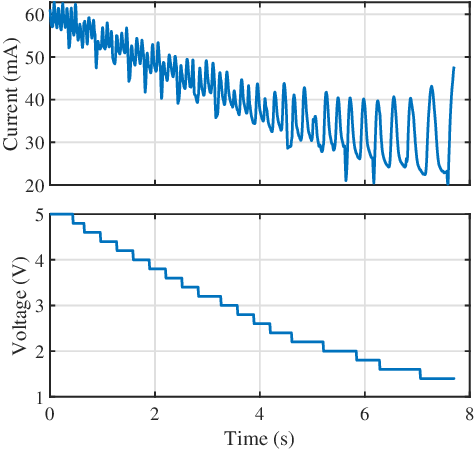} % 單欄圖片
  \caption{As the motor driving voltage decreases, the magnitude of the current rise during the loaded phase increases.}
  \label{fig:fig1} % 用於在文本中引用圖片的標籤
\end{figure}

Battery-powered micro actuators are fundamental building blocks in emerging biomedical devices, including ingestible electronics\cite{steiger2019ingestible}, minimally invasive surgical tools, wearable systems, and implantable platforms\cite{Sun2024MicroNanoMedicine, Yan2022ImplantableActuators}. In these applications, micro DC brushed motors remain attractive due to their compact size, simplicity, and compatibility with low-voltage embedded systems. However, their operation is severely constrained by limited battery capacity\cite{ben2015power}, strict thermal limits\cite{zangooei2021thermal,shepherd2018development}, and highly variable physiological loading conditions\cite{iacovacci2024medical}, making energy-efficient actuation a critical challenge.

\par
% Previous studies on micro DC motors have mainly focused on motor efficiency\cite{Sousa2017InductionMotorEfficiency}, torque improvement, mechanical structure design\cite{Barkas2020BrushedDCReview, Kuczmann2024DCModelingControl}, and condition monitoring based on electrical signals\cite{Niu2023CurrentSignatureReview, Krause2023NonintrusiveMCSA,Henao2014FaultDiagnosisReview}. Although these studies are helpful for improving motor performance and reliability, they have paid less attention to the energy consumption of motors. Under energy-constrained conditions, reducing power consumption may not be sufficient, since the energy required to complete one mechanical cycle is also affected by the cycle duration. As the driving voltage decreases, the motor power consumption may decrease, but the motor speed may also decrease, resulting in a longer mechanical cycle. Therefore, a lower driving voltage does not necessarily lead to lower energy consumption per cycle.
In practice, micro motors in biomedical environments are often subject to uncertain and time-varying mechanical loads, arising from interactions with soft tissues, fluids, or dynamically changing anatomical structures\cite{cao2024robotic}. To ensure reliable operation and avoid stalling under such conditions, these motors are typically driven using conservative voltage or current margins. While this approach guarantees sufficient torque delivery, it introduces a fundamental inefficiency: under nominal or light loading, the motor is operated with unnecessarily high current and speed, leading to excessive energy consumption, rapid battery depletion, and elevated thermal dissipation, which can compromise both device longevity and patient safety.
\par

% In many motor-driven applications, the mechanical load is not uniformly distributed over the entire motion cycle. Instead, the motor may experience a higher load during only a specific portion of the cycle, for example when driving a linkage mechanism or compressing an elastic element. Under such conditions, loading variation is often reflected as a localized change in the motor current waveform rather than a regular shift over the full cycle. In the mechanism considered in this work, this effect appears during the spring-compression phase, where the motor is subjected to a higher mechanical load and the current waveform exhibits a distinct current rise. As illustrated in Fig.~\ref{fig:fig1}, when the driving voltage is reduced, the current rise during the loaded phase becomes more pronounced. This observation suggests that the waveform shape of the elevated-current segment may provide useful information for characterizing loading variation, which motivates the waveform-based load metric developed in this work.
From a mechatronic perspective, this inefficiency reflects the absence of closed-loop adaptation between electrical actuation and physiological loading. Although the electromechanical behavior of DC motors is well understood\cite{barkas2020brushed}, including the existence of load-dependent operating points that minimize energy consumption, such knowledge is rarely exploited in real time within biomedical systems. Instead, actuation strategies are typically fixed or pre-calibrated, lacking the ability to respond dynamically to changing in vivo conditions.

\par
Recent advances in embedded sensing and model-based estimation enable a new approach to this problem. By leveraging readily available signals—such as motor voltage, current, and rotational speed—it is possible to infer the external load experienced by the motor and adjust its operating conditions accordingly\cite{vazquez2011new,linares2010load}. This creates an opportunity for self-adaptive, energy-aware actuation, in which the motor continuously aligns its operating point with the instantaneous mechanical demand while respecting safety constraints.
\par
% Motivated by this need, this paper proposes a model-based adaptive voltage control framework for battery-powered micro DC brushed motors in biomedical applications. The core idea is to couple a lightweight load torque estimation mechanism with an adaptive voltage regulation strategy, enabling the motor to automatically converge toward its minimum-energy operating condition under unknown and time-varying loads. Unlike conventional approaches that rely on conservative overvoltage to ensure robustness, the proposed method achieves both reliable operation and energy efficiency by explicitly accounting for motor loss mechanisms and load-dependent optimality.

Motivated by this need, this paper proposes an adaptive voltage control framework for battery-powered micro DC brushed motors in biomedical applications. The core idea is to couple a load metric mechanism with an adaptive voltage regulation strategy, enabling the motor to automatically converge toward its minimum-energy operating condition under unknown and time-varying loads. The load metric was developed based on the observation that the current rise during the loaded phase becomes more pronounced at lower driving voltages, as shown in Fig.~\ref{fig:fig1}. Unlike conventional approaches that rely on conservative overvoltage to ensure robustness, the proposed method achieves both reliable operation and energy efficiency by explicitly accounting for motor loss mechanisms and load-dependent optimality.

\par
The proposed approach is particularly suited for biomedical systems due to its minimal hardware requirements and inherent safety awareness. It requires only standard electrical and kinematic measurements, making it compatible with highly constrained embedded platforms. Furthermore, the same model used for energy optimization naturally supports overload detection and thermal-aware operation, which are essential for safe interaction with biological tissues.

% To address this problem, this paper proposes a minimum-energy operating-point tracking method for a brushed DC motor under varying loading conditions. First, the relationships among driving voltage, motor speed, power consumption, and energy consumption per cycle are analyzed to examine the existence of a minimum-energy operating point. Then, a load-sensitive metric is developed from the motor current waveform to reflect load-induced changes during each mechanical cycle. Based on this metric, an online tracking method is designed to detect loading variation and adjust the driving voltage toward the new minimum-energy operating region. Finally, the proposed method is experimentally evaluated in terms of load discrimination, response time, and convergence voltage, in order to verify its effectiveness for real-time energy-aware operation.
% To address this problem, this paper proposes a real-time minimum-energy operating-point tracking method for a brushed DC motor under varying loading conditions. 
The main contributions of this work are as follows.
1) The relationship between driving voltage and cycle energy is experimentally characterized under different loading conditions, revealing the existence and load dependence of a minimum-energy operating point.
2) A computationally simple load-sensitive metric is developed from the motor current waveform to capture load-induced changes during each mechanical cycle.
3) A real-time tracking method is designed to detect load variation and adjust the driving voltage toward the new minimum-energy operating region.
% The proposed method is validated through hardware experiments in terms of loading variations, response time, and convergence voltage, demonstrating its effectiveness for real-time energy-aware operation.
By enabling actuators to adapt to their physiological environment in real time, this work advances the development of intelligent, energy-efficient biomedical mechatronic systems capable of prolonged and safe operation within the human body.

\section{EXPERIMENTAL SETUP AND MODEL}

\subsection{Experimental Setup}

\subsubsection{System architecture}
\begin{figure}[!tbp] % 或者 [!htb], [!h] 等，根據您的排版需求調整
  \centering
  \includegraphics[width=\columnwidth]{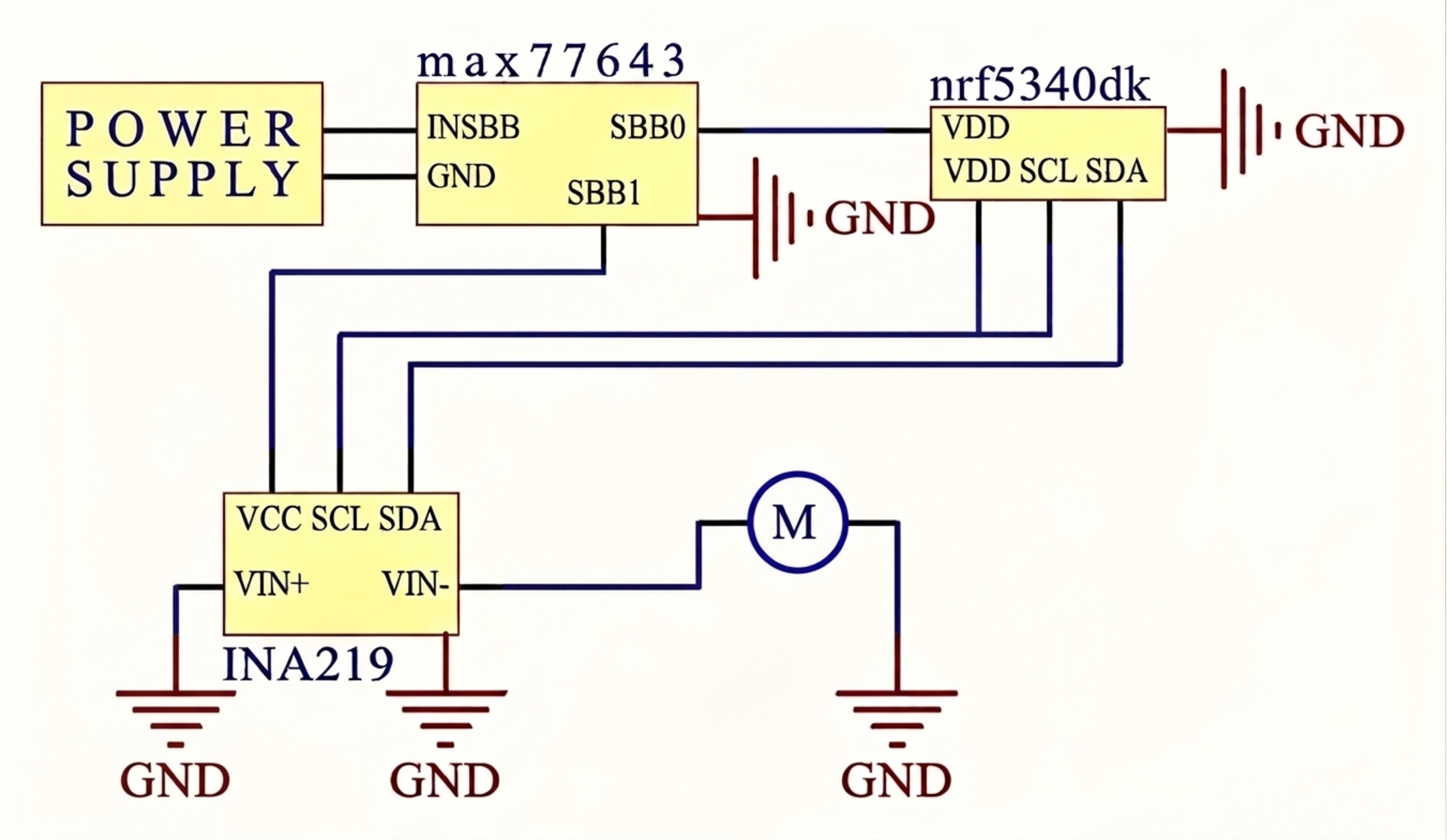} % 單欄圖片
  % 或者
  % \includegraphics[width=0.8\columnwidth]{figure.eps} % 圖片寬度為單欄的80%
  % 或者
  % \includegraphics[scale=0.5]{figure.eps} % 按比例縮放圖片
  \caption{The circuit schematic of the setup}
  \label{fig:circuit} % 用於在文本中引用圖片的標籤
\end{figure}

A DC power supply (UNI-T UTP1306S) is employed as the power source in the experimental system. A micro brushed DC motor equipped with a three-stage planetary gearbox with an overall gear ratio of 136:1 is employed as the actuator. A programmable output voltage DC-DC converter (Maxim Integrated, MAX77643) is employed. The motor current is measured through a current monitor (Texas Instruments, INA219) at a sampling rate of 500Hz. 
% The rotational period of the motor is determined using a 9-axis motion tracking device with a magnetometer (TDK InvenSense, ICM-20948); specifically, the period is calculated by detecting periodic magnetic field variations generated by a small magnet fixed to the motor shaft during rotation. 
The microcontroller unit (Nordic Semiconductor, nRF5340 DK) is employed to read sensor data and adjust the DC-DC converter output voltage with $I^2C$ interface. 
The overall system architecture is as shown in  Fig.~\ref{fig:circuit}.

\subsubsection{Testing mechanism and load configuration}

We designed and constructed a test mechanism to apply a controllable mechanical load to the motor.
% , as shown in \Cref{fig:mechanism}. 
The mechanism converts the rotational motion of motor shaft into linear motion. The mechanical component compresses a spring during the rotation of the motor shaft thereby generating a reaction force to the motor. Springs with different spring constants are used to test and analyze the response characteristics of the motor current under different load conditions.
% \begin{figure}[!tbp]
%   \centering
%   % 第一張子圖
%   \begin{subfigure}[b]{0.48\columnwidth} % [b] 表示底部對齊，寬度約為單欄的一半
%     \centering
%     \includegraphics[width=\linewidth]{figure/mechanism_real.eps} % 圖片寬度充滿subfigure
%     \caption{}
%     \label{fig:subfig1}
%   \end{subfigure}
%   \hfill % 在子圖之間添加一些水平間距，使其分開
%   % 第二張子圖
%   \begin{subfigure}[b]{0.48\columnwidth} % [b] 表示底部對齊，寬度約為單欄的一半
%     \centering
%     \includegraphics[width=\linewidth]{figure/mechanism_spring.eps} % 圖片寬度充滿subfigure
%     \caption{}
%     \label{fig:subfig2}
%   \end{subfigure}
%   \caption{Testing mechanism: (a) The actual one, (b)The designed one}
%   \label{fig:mechanism}
% \end{figure}

\subsection{Brushed DC motor Dynamic Characterization}

A simplified electrical equation\cite{Verstraten2016ModelingDesignGearedDC} of the brushed DC motor can be described by:
\begin{equation}
    V_m = I_m \cdot R_m + L_m \cdot \frac{dI_m}{dt} + V_{emf}
\end{equation}
where $V_m$ is the driving voltage, $I_m$ is the motor current, 
$R_m$ is the winding resistance, $L_m$ is the winding inductance, 
and $V_{emf}= K_e \cdot \omega_m$ is the back electromotive force (EMF) proportional 
to the shaft angular velocity $\omega_m$
where $K_e$ is the motor back-EMF constant.

The mechanical equation of the motor shaft can be expressed as:
\begin{equation}
    \tau_m = J_m \cdot \frac{d\omega_m}{dt} + b_m \cdot \omega_m + \tau_{\text{load}}
\end{equation}
where $\tau_m = K_t \cdot I_m$ is the output shaft torque of a brushed DC motor, $K_t$ is the motor torque constant, $J_m$ is the rotor inertia, 
$b_m$ is the viscous drag coefficient, and $\tau_{\text{load}}$ is the load torque.

When the motor reaches the steady state, the current variation and the angular acceleration term become negligible such that
$\frac{dI_m}{dt} \approx 0$ and $\frac{d\omega_m}{dt} \approx 0$. Thus, the motor current can be approximated as:
\begin{equation}
    I_m \approx \frac{V_m - K_e \cdot \omega_m}{R_m}
    \label{eq:Iss}
\end{equation}
% This equation indicates that the motor current is determined by the difference between the applied voltage and the back-EMF generated by the motor speed.

% When the motor rotation speed is under steady state, the angular acceleration term is negligible where $\frac{d\omega_m}{dt} \approx 0$. 
The rotation speed can be derived from the mechanical equation as:
\begin{equation}
    \omega_m = \frac{\frac{K_t}{R_m} \cdot V_m - \tau_{\text{load}}}{b_m + \frac{K_t \cdot K_e}{R_m}}
    \label{eq:speed}
\end{equation}
Eq.~\eqref{eq:Iss} and ~\eqref{eq:speed} indicate that the motor current is determined by the difference between the applied voltage and the back-EMF generated by the motor speed and the motor speed $\omega_m$ decreases as the driving voltage $V_m$ decreases.

The power consumption of the motor\cite{huang2023power} can be calculated as:
\begin{equation}
    P_m(t) = V_m(t) \cdot I_m(t)
    \label{power}
\end{equation}
The energy consumption per cycle\cite{Verstraten2016EnergyConsumptionGearedDC} can be calculated as:
\begin{equation}
    E_m = \int_{t_0}^{t_0 + T_s} P_m(t) dt = \int_{t_0}^{t_0 + T_s} V_m(t) \cdot I_m(t) dt
    \label{energy}
\end{equation}
where $T_s$ is the period of one revolution.

% ---------------------------------------------------------------
% III. Method
% ---------------------------------------------------------------

\section{Method}
In this work, we considered the relationship between voltage and power/energy consumption under different loadings and defined a load metric which can detect the loading variation. We then constructed the minimum-energy operating-point tracking algorithm that dynamically adjusts the driving voltage based on the motor current.

\subsection{Definition of The Energy Consumption Per Cycle}
Based on the motor model, the continuous-time power consumption and energy consumption per cycle are defined as shown in Eq. ~\eqref{power} and ~\eqref{energy}. However, the voltage and current signals are acquired in discrete form. Besides, one complete motor cycle corresponds to the energy required for one action. Therefore, a discrete-time energy consumption per cycle estimate is introduced as:
\begin{equation}
    E_{\text{m}} = \frac{1}{N} \sum_{k=1}^{N} V_m[k] \cdot I_m[k] \cdot T_s
    \label{eq:energy_disc}
\end{equation}
% where $V_m[k]$ and $I_m[k]$ denote the sampled voltage and current at time step $k$ and $T_s$ is the period time of one cycle which is calculated from magnetic field data.
where $V_m[k]$ and $I_m[k]$ denote the sampled voltage and current
at time step~$k$, $N$ is the total number of samples within one
mechanical cycle, and $T_s$ is the period of that cycle calculated
from magnetic field data. Since the driving voltage is held constant
during each measurement interval, $V_m[k] = V_m$ for all~$k$. The
term $\frac{1}{N}\sum I_m[k]$ represents the average motor current over
one cycle, so the product $V_m \cdot \frac{1}{N}\sum I_m[k] \cdot T_s$
corresponds to the average power multiplied by the cycle duration,
giving the energy consumption per cycle.
\par
Since the energy estimation of a single cycle is more sensitive to cycle-to-cycle fluctuations and measurement noise, the energy is further averaged over 1, 3, and 5 mechanical cycles. This averaging process improves the robustness of the energy estimate, but it also increases the time required to update the operating point. Therefore, different averaging-cycle numbers are considered to balance response speed and estimation stability.

% \subsection{Minimum-Energy Voltage Search Strategy}
% The voltage search starts at a safe initial voltage
% $V_{\mathrm{init}} = 5$\,V, ensuring sufficient torque margin. The voltage
% is then decremented by a fixed step $\Delta V = 0.2$\,V at each iteration.
% At each voltage level, the motor operates for $N_c$ consecutive revolutions
% ($N_c \in \{1, 3, 5\}$). The cycle energy is estimated using
% \eqref{eq:energy_disc} and averaged over $N_c$ cycles. The search continues
% until the estimated energy begins to increase, and the voltage at which the
% minimum averaged energy was recorded is selected as the optimal operating
% point:
% \begin{equation}
%   V^* = \arg\min_{V} \bar{E}_m(V).
%   \label{eq:Vstar}
% \end{equation}

\subsection{Load Metric for Detecting Loading Variation}
In the proposed mechanism, the motor current rises during the spring-compression phase, where the motor experiences elevated mechanical loading. As the load increases, the current waveform typically exhibits not only a larger current excursion but also a longer high-current interval within one mechanical cycle. Therefore, a load-sensitive metric should capture both the magnitude and the duration of this phenomenon.
\par
\begin{figure}[t]
  \centering
  \includegraphics[width=0.9\columnwidth]{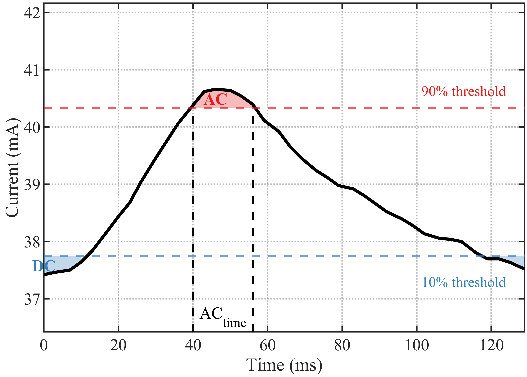}
  \caption{Definition of the AC component, DC component, and AC time from
           the motor current waveform.}
  \label{fig:fig3}
\end{figure}

Peak current alone is highly sensitive to noise, while the average current over a full cycle tends to obscure the information associated with the loaded phase. In contrast, the proposed metric preserves the features caused by load variation, namely the amplitude of the current rise and the duration of the elevated-current interval.
To maintain computational simplicity, the current waveform is characterized by three quantities: the high-current component $AC$, the baseline component $DC$, and the duration of the high-current phase $AC_{time}$. These waveform features are illustrated in Fig.~\ref{fig:fig3}.

The AC and DC components are extracted using adaptive threshold-based
statistics over one mechanical cycle of
$N_s = f_{\mathrm{sample}}/f_{rev}$ samples where $f_{\mathrm{sample}}$ is the sampling frequency of the sensor and $f_{rev}$ is the frequency of the motor revolution:
\begin{equation}
  AC_{\mathrm{thr}} = I_{\mathrm{valley}}
    + 0.9 \times (I_{\mathrm{peak}} - I_{\mathrm{valley}}),
  \label{eq:ACthr}
\end{equation}
\begin{equation}
  AC = \mathrm{mean}\!\left(I_m[k] \mid
        I_m[k] > AC_{\mathrm{thr}},\;
        k = 1,\ldots,N_s \right),
  \label{eq:AC}
\end{equation}
\begin{equation}
  DC_{\mathrm{thr}} = I_{\mathrm{valley}}
    + 0.1 \times (I_{\mathrm{peak}} - I_{\mathrm{valley}}),
  \label{eq:DCthr}
\end{equation}
\begin{equation}
  DC = \mathrm{mean}\!\left(I_m[k] \mid
        I_m[k] < DC_{\mathrm{thr}},\;
        k = 1,\ldots,N_s \right).
  \label{eq:DC}
\end{equation}
\par
Here, AC and DC do not denote alternating and direct current; they refer to the elevated-current and baseline-current components extracted from one mechanical cycle.
\par
The AC time is the duration of the high-current phase per cycle:
\begin{equation}
  \begin{split}
    AC_{\mathrm{time}} = t_{AC_{\mathrm{end}}} - t_{AC_{\mathrm{start}}}, 
    \text{where } I_m[k] > AC_{\mathrm{thr}}, \\
    t \in [t_{AC_{\mathrm{start}}}, t_{AC_{\mathrm{end}}}].
  \end{split}
  \label{eq:ACtime}
\end{equation}

For this reason, the proposed metric combines the current-rise magnitude $(AC-DC)$ and the high-current duration $AC_{time}$ and is defined as:
\begin{equation}
  \mathcal{L} = (AC - DC) \times AC_{\mathrm{time}}.
  \label{eq:metric}
\end{equation}
The product of the $(AC-DC)$ and $AC_{\mathrm{time}}$ is to amplify both the increased amplitude and extended duration characteristics.
Also, this formulation preserves the waveform features associated with load variation while remaining simple for real-time implementation.

\subsection{The Minimum-Energy Operating-Point Tracking Algorithm}
To identify the minimum-energy consumption and detect load variation, the Minimum-Energy Operating-Point Tracking Algorithm consists of two phases, minimum-energy voltage search strategy and online load variation detection.
\par
In Phase I, the voltage search starts at a safe initial voltage
$V_{\mathrm{init}} = 5$\,V in the voltage search range $[V_{\min},\, V_{\max}]$, ensuring sufficient torque margin. The voltage
is then decremented by a fixed step $\Delta V = 0.2$\,V at each iteration.
At each voltage level, the motor operates for $N_c$ consecutive revolutions
($N_c \in \{1, 3, 5\}$). The energy consumption per cycle is estimated using
\eqref{eq:energy_disc} and averaged over $N_c$ cycles. The search continues
until the estimated energy begins to increase, and the voltage at which the
minimum averaged energy was recorded is selected as the optimal operating
point:
\begin{equation}
  V^* = \arg\min_{V} \bar{E}_m(V).
  \label{eq:Vstar}
\end{equation}
\par
In Phase II, the load metric $\mathcal{L}$
is monitored online to adapt the operating voltage to load variations with two load-change thresholds $\Delta_+$ and $\Delta_-$
for detecting load increases and decreases respectively. A detected load decrease triggers a direct downward search from the current operating point, whereas a detected load increase first raises the voltage until $\mathcal{L}$ converges with a convergence tolerance $\epsilon_{\text{Load}}$ and then performs another downward search.

% ---------------------------------------------------------------
% pseudo code
% ---------------------------------------------------------------

\begin{algorithm}[!htbp]
\small
\caption{Minimum-Energy Operating-Point Tracking}
\label{alg:meopt}
\begin{algorithmic}[1]
\Require $V_{\max}$, $V_{\min}$, $\Delta V$, $\Delta_+$, $\Delta_-$, $\epsilon_{\text{Load}}$
\Ensure Optimal operating voltage $V^*$
 
\Statex \textit{// Phase~I: Initial energy search}
\State $V \gets V_{\max}$; \Call{SetVoltage}{$V$}
\State $E_{\min} \gets$ \Call{MeasureEnergy}{$V, I_m, \omega_m$}; \quad $V^* \gets V$
\While{$V - \Delta V \geq V_{\min}$}
    \State $V \gets V - \Delta V$; \Call{SetVoltage}{$V$}
    \State $E \gets$ \Call{MeasureEnergy}{$V, I_m, \omega_m$}
    \If{$E \leq E_{\min}$}
        \State $E_{\min} \gets E$; \quad $V^* \gets V$
    \Else
        \State \Call{SetVoltage}{$V^*$}; \textbf{break}
    \EndIf
\EndWhile
 
\Statex
\Statex \textit{// Phase~II: Load-change monitoring at $V^*$}
\State $\mathcal{L}_{\text{ref}} \gets$ \Call{MeasureLoadMetric}{$V^*$}
\While{system is running}
    \State $\mathcal{L} \gets$ \Call{MeasureLoadMetric}{$V^*$}
    \State $\delta \gets \mathcal{L} - \mathcal{L}_{\text{ref}}$
 
    \If{$\delta < \Delta_-$} \Comment{high $\to$ low load}
        \State $E_{\min} \gets$ \Call{MeasureEnergy}{$V^*, I_m, \omega_m$}; \quad $V \gets V^*$
        \While{$V - \Delta V \geq V_{\min}$}
            \State $V \gets V - \Delta V$; \Call{SetVoltage}{$V$}
            \State $E \gets$ \Call{MeasureEnergy}{$V, I_m, \omega_m$}
            \If{$E \leq E_{\min}$}
                \State $E_{\min} \gets E$; \quad $V^* \gets V$
            \Else
                \State \Call{SetVoltage}{$V^*$}; \textbf{break}
            \EndIf
        \EndWhile
        \State $\mathcal{L}_{\text{ref}} \gets$ \Call{MeasureLoadMetric}{$V^*$}
 
    \ElsIf{$\delta > \Delta_+$} \Comment{low $\to$ high load}
        \Statex \hspace{\algorithmicindent}\hspace{\algorithmicindent}\textit{// Step~2a: Raise $V$ until load metric converges}
        \State $\mathcal{L}_{\text{prev}} \gets \mathcal{L}$; \quad $V \gets V^*$
        \While{$V + \Delta V \leq V_{\max}$}
            \State $V \gets V + \Delta V$; \Call{SetVoltage}{$V$}
            \State $\mathcal{L} \gets$ \Call{MeasureLoadMetric}{$V$}
            \If{$|\mathcal{L} - \mathcal{L}_{\text{prev}}| < \epsilon_{\text{Load}}$}
                \State \textbf{break}
            \EndIf
            \State $\mathcal{L}_{\text{prev}} \gets \mathcal{L}$
        \EndWhile
        \Statex \hspace{\algorithmicindent}\hspace{\algorithmicindent}\textit{// Step~2b: Downward energy sweep from converged $V$}
        \State $E_{\min} \gets$ \Call{MeasureEnergy}{$V, I_m, \omega_m$}; \quad $V^* \gets V$
        \While{$V - \Delta V \geq V_{\min}$}
            \State $V \gets V - \Delta V$; \Call{SetVoltage}{$V$}
            \State $E \gets$ \Call{MeasureEnergy}{$V, I_m, \omega_m$}
            \If{$E \leq E_{\min}$}
                \State $E_{\min} \gets E$; \quad $V^* \gets V$
            \Else
                \State \Call{SetVoltage}{$V^*$}; \textbf{break}
            \EndIf
        \EndWhile
        \State $\mathcal{L}_{\text{ref}} \gets$ \Call{MeasureLoadMetric}{$V^*$}
    \EndIf
\EndWhile
\end{algorithmic}
\end{algorithm}

% ---------------------------------------------------------------
% IIII. RESULTS AND DISCUSSION
% ---------------------------------------------------------------

\section{RESULTS AND DISCUSSION}

\subsection{The Relationship between Driving Voltage, Power Consumption and Motor Speed}

\begin{figure}[t]
  \centering
  \subfloat[]{\includegraphics[width=0.5\columnwidth]{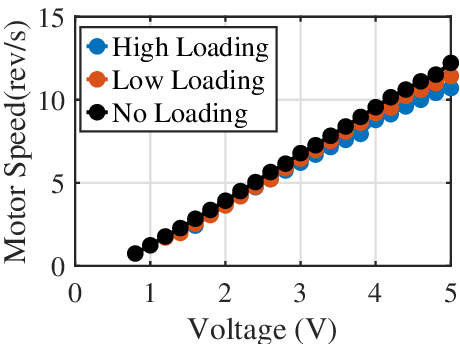}\label{fig:fig4a}}
  \hfill
  \subfloat[]{\includegraphics[width=0.5\columnwidth]{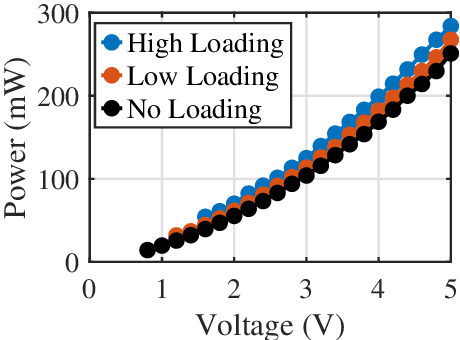}\label{fig:fig4b}}
  \caption{(a)Motor speed with varying voltages under different load conditions, (b)Power consumption with varying voltages under different load conditions.}
  \label{fig:4a4b}
\end{figure}

To first examine the mechanical response of the motor under different driving voltages and loading conditions, the motor speed is analyzed.
Fig.~\ref{fig:fig4a} illustrates the relationship between the driving voltage and the motor speed under different loading conditions. 
As the driving voltage decreases, the motor speed decreases accordingly, which is consistent with the voltage–speed relationship described in Eq.~\eqref{eq:speed}. In addition, at the same driving voltage, the speed under high loading is lower than that under low loading.

To further examine the corresponding electrical behavior, the motor power consumption is then analyzed under the same conditions.
Fig.~\ref{fig:fig4b} shows that the motor power consumption decreases with decreasing driving voltage under different loadings. As the driving voltage decreases, the motor power consumption also decreases, which agrees with Eq.~\eqref{power}, since both the applied voltage and the motor current tend to decrease with a lower driving voltage.

However, a lower driving voltage not only reduces the power consumption but also slows down the motor, thereby increasing the duration of each mechanical cycle. This indicates that a reduction in power does not necessarily imply a reduction in the energy required to complete one cycle. In other words, there exists a trade-off between lower power consumption and longer operation time.
To identify the voltage at which the energy consumption per cycle reaches a minimum under each loading condition, the energy per cycle is further examined in the next subsection.

\subsection{The Minimum-energy Voltage Exists and Shifts with Loading}

\begin{figure}[t]
  \centering
  \includegraphics[width=0.8\columnwidth]{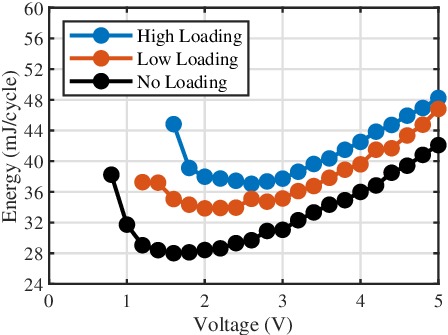}
  \caption{Energy consumption per cycle with varying voltages under different loading conditions.}
  \label{fig:fig5}
\end{figure}

Based on the trade-off identified in Section IV-A between reduced power consumption and prolonged cycle duration, Fig. ~\ref{fig:fig5} further examines the energy consumption per cycle under different driving voltages and loading conditions. As shown in Fig.  ~\ref{fig:fig5}, the energy consumption per cycle exhibits a non-monotonic relationship with driving voltage, decreasing initially and then rising again at lower voltages, thereby forming a distinct minimum point. The minimum energy occurs at approximately 2V under low loading, corresponding to about 33.9 mJ/cycle.
\par
In addition, the voltage corresponding to the minimum energy point shifts toward a higher value as the loading increases. Under no-loading condition, the minimum energy occurs at a relatively low voltage, whereas under high loading, the motor requires a higher voltage to reach the energy-optimal point. The minimum-energy voltage shifts from approximately 1.6V under no loading to 2.6V under high loading. This is because a heavier load demands more torque and further reduces the motor speed at any given voltage, causing the cycle duration to increase more rapidly at lower voltages.
\par

These results yield two main observations. First, each loading condition is associated with an energy-optimal driving voltage that minimizes the energy consumption per cycle. Second, the energy-optimal voltage shifts with the load. Hence, a constant driving voltage is generally insufficient to maintain minimum energy consumption under load variations. This observation motivates the load-sensitive metric introduced in the following subsection for load variation detection and adaptive voltage regulation.

\subsection{Load Metric}

% \begin{figure}[t]
%   \centering
%   \includegraphics[width=0.9\columnwidth]{figure/fig6.eps}
%   \caption{Load metric with varying voltages under different loading conditions.}
%   \label{fig:fig6}
% \end{figure}
Because the minimum-energy operating voltage is loading-dependent, adaptive voltage regulation requires a metric that can reliably indicate load variation. As shown in Fig. 6, the proposed load metric is consistently higher in the high-loading condition than in the low-loading condition over the examined voltage range. This separation between the two curves demonstrates that the metric has sufficient sensitivity to distinguish different loading states.

Moreover, the turning region of the metric appears near the minimum-energy voltage observed in Fig. 5. This result suggests that the proposed metric carries useful information about the operating region surrounding the energy-optimal point. Although the metric is not used to replace direct energy evaluation, it provides a lightweight current-domain indicator of load-induced operating-point shift.

Therefore, the metric is suitable for online feedback in the proposed control framework, enabling the system to detect loading variation and guide the search toward the new minimum-energy operating point after a load variation.

% Since the minimum-energy voltage depends on the loading condition, a load metric is required to support adaptive voltage regulation. Fig.~\ref{fig:fig6} shows the proposed metric under different driving voltages for low-loading and high-loading conditions. It can be observed that the metric under high loading is consistently higher than that under low loading, indicating that the proposed metric can effectively distinguish different loading conditions. Moreover, the turning point of the metric curve appears near the minimum-energy voltage observed in Fig.~\ref{fig:fig5}. This suggests that the proposed metric is not only sensitive to load variations but also informative of the operating region near the minimum-energy point. Therefore, it can serve as effective feedback for guiding the minimum-energy voltage search toward the new minimum-energy point after load variation.
\begin{figure}[t]
  \centering
  \includegraphics[width=0.8\columnwidth]{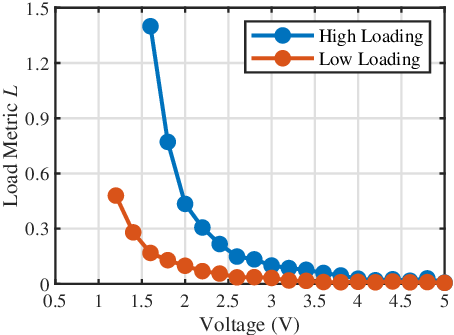}
  \caption{Load metric with varying voltages under different loading conditions.}
  \label{fig:fig6}
\end{figure}

\subsection{Response Time after Loading Variations}

To evaluate the real-time adaptability of the proposed method, the response time after load transitions was measured for both low-to-high and high-to-low loading changes. Here, the response time is defined as the time required for the method to reach the new minimum-energy operating point after the loading condition changes, assuming that the motor is initially operating at the minimum-energy point of the original load.
\par
Figs. 7(a) and 7(b) compare the response time obtained using 1-cycle, 3-cycle, and 5-cycle averaging under load-increase and load-decrease conditions, respectively. In both transition directions, the response time increases as the number of averaging cycles increases. This trend is consistent with the energy-estimation strategy described in Section III, where averaging over more mechanical cycles improves robustness against cycle-to-cycle fluctuation and measurement noise, but also delays the operating-point update.
\par
Among the three settings, 1-cycle averaging achieves the fastest adaptation, whereas 5-cycle averaging yields the slowest response. However, although fewer averaging cycles improve adaptation speed, they are also more susceptible to short-term fluctuation in the cycle-energy estimate. In contrast, using more averaging cycles improves the robustness of the update decision at the cost of slower response.
\par
The consistent trend observed in both load-increase and load-decrease transitions indicates that the effect of cycle averaging on adaptation speed is stable across different operating-point recovery scenarios. Considering the trade-off between response speed and estimation robustness, the 3-cycle setting provides a practical compromise for real-time implementation.

\begin{figure}[t]
  \centering
  \subfloat[Low-to-high loading]{\includegraphics[width=0.5\columnwidth]{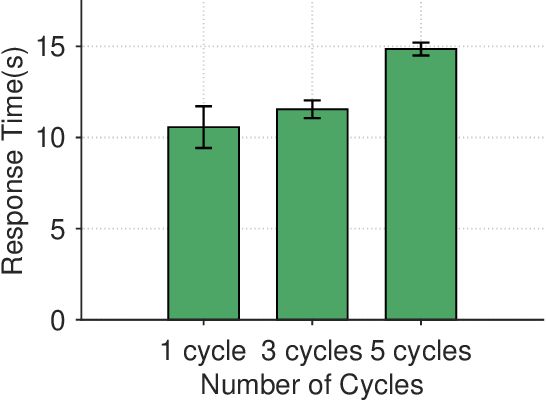}\label{fig:fig7a}}
  \hfill
  \subfloat[High-to-low loading]{\includegraphics[width=0.5\columnwidth]{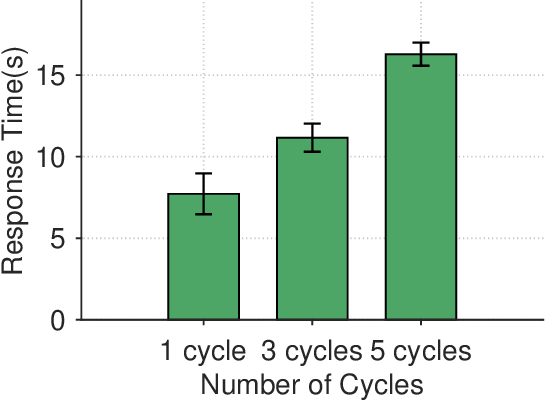}\label{fig:fig7b}}
  \caption{Response time for different numbers of averaging cycles under load-increase and load-decrease conditions.}
  \label{fig:response_time}
\end{figure}

\subsection{Convergence Voltage after Loading Variations}

\begin{figure}[t]
  \centering
  \subfloat[Low-to-high loading]{\includegraphics[width=0.48\columnwidth]{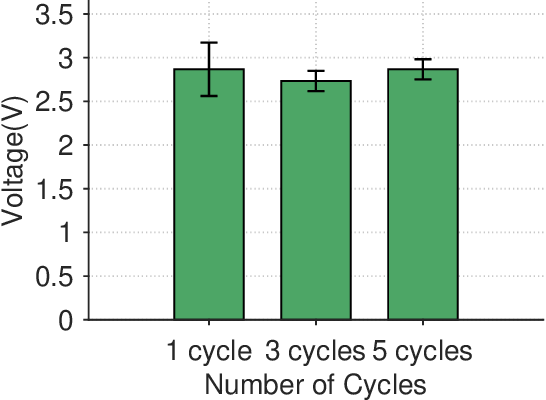}\label{fig:fig8a}}
  \hfill
  \subfloat[High-to-low loading]{\includegraphics[width=0.48\columnwidth]{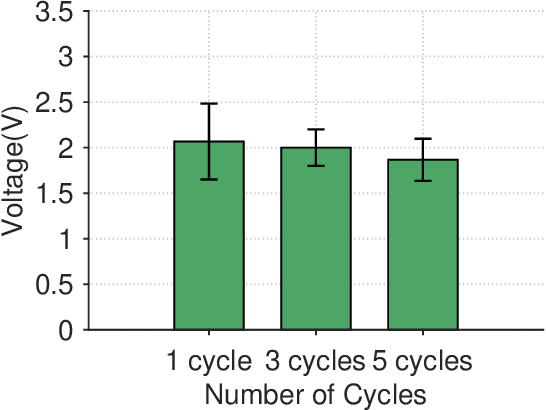}\label{fig:fig8b}}
  \caption{Convergence voltage for different numbers of averaging cycles under load-increase and load-decrease conditions.}
  \label{fig:convergence_voltage}
\end{figure}

While the response time characterizes how quickly the proposed method reacts to a load variation, the convergence voltage reflects how accurately and consistently the new operating point is reached after the loading condition changes. Figs.~\ref{fig:convergence_voltage} show the convergence voltages obtained using 1-cycle, 3-cycle, and 5-cycle averaging for low-to-high and high-to-low loading variations, respectively.
\par
In both transition directions, the convergence voltages obtained under all three averaging settings fall within the voltage region associated with the new minimum-energy operating point. This indicates that the proposed method can reliably guide the motor toward the energy-optimal operating region after the load changes, rather than converging to fixed voltage.
\par
However, the error bars show that the repeatability of the final voltage depends on the number of averaging cycles. Using fewer averaging cycles leads to larger variation in the convergence voltage, since the operating-point update becomes more sensitive to cycle-to-cycle fluctuation and measurement noise. In contrast, increasing the number of averaging cycles improves the consistency of the final voltage, although, as shown in Fig.~\ref{fig:response_time}, this improvement comes at the cost of slower response.
\par
Taken together, these results indicate that 3-cycle averaging provides a practical compromise between convergence consistency and adaptation speed. It maintains better final-voltage repeatability than the 1-cycle case while avoiding the slower response associated with 5-cycle averaging, making it suitable for real-time minimum-energy operating-point tracking.
\par

% ---------------------------------------------------------------
% 5. Conclusions
% ---------------------------------------------------------------

\section{CONCLUSIONS AND FUTURE WORK}

This paper presented a minimum-energy operating-point tracking method for a brushed DC motor under varying loading conditions. Instead of evaluating only power consumption, the proposed approach considered the energy consumption per mechanical cycle, which is more directly related to the energy required to complete one actuation. Experimental results showed that the energy consumption per cycle has a non-monotonic relationship with driving voltage and exhibits a minimum point, indicating that reducing the driving voltage does not always minimize the energy consumption. In addition, the voltage corresponding to the minimum-energy point shifts with the loading condition, showing that a fixed driving voltage cannot guarantee minimum-energy operation under load variation.
\par
To address this issue, a load metric derived from the motor current waveform was introduced to detect load variation and guide the voltage update process. The proposed method was validated through load-transition experiments using response time and convergence voltage. The results showed that the method can effectively guide the motor toward the new minimum-energy operating region after load changes, while the 3-cycle averaging setting provides a practical balance between adaptation speed and convergence consistency. These findings suggest that the proposed framework is suitable for energy-constrained miniature motor-driven systems where extending operational lifetime is important.
\par
% Future work will investigate how different gearbox configurations and voltage step sizes affect the energy–voltage curve, including the location of the minimum-energy point, as well as the response speed and convergence behavior of the proposed tracking method.

Future work will focus on quantifying how gearbox ratio influences the energy–voltage curve and the location of the minimum-energy point. We hypothesize that higher gearbox reduction ratios may shift the minimum-energy point toward a lower voltage region. In addition, the effects of different voltage step sizes on the convergence speed, response behavior, and operating-point accuracy of the proposed tracking method will be systematically analyzed.
\addtolength{\textheight}{-12cm}   % This command serves to balance the column lengths

%%%%%%%%%%%%%%%%%%%%%%%%%%%%%%%%%%%%%%%%%%%%%%%%%%%%%%%%%%%%%%%%%%%%%%%%%%%%%%%%

%%%%%%%%%%%%%%%%%%%%%%%%%%%%%%%%%%%%%%%%%%%%%%%%%%%%%%%%%%%%%%%%%%%%%%%%%%%%%%%%

%%%%%%%%%%%%%%%%%%%%%%%%%%%%%%%%%%%%%%%%%%%%%%%%%%%%%%%%%%%%%%%%%%%%%%%%%%%%%%%%

%%%%%%%%%%%%%%%%%%%%%%%%%%%%%%%%%%%%%%%%%%%%%%%%%%%%%%%%%%%%%%%%%%%%%%%%%%%%%%%%

% References are important to the reader; therefore, each citation must be complete and correct. If at all possible, references should be commonly available publications.
% \nocite{*}
% \onecolumn
\bibliographystyle{IEEEtran}
\bibliography{references}

\end{document}